\documentclass[prl,tightenlines,
twocolumn,
nofootinbib]{revtex4}

\usepackage{graphicx}  
\usepackage{bm}  
\usepackage{amsmath}
\usepackage{epsf}
\newcommand{\beq}{\begin{equation}}
\newcommand{\eeq}{\end{equation}}
\newcommand{\bea}{\begin{eqnarray}}
\newcommand{\eea}{\end{eqnarray}}

\def\vec#1{{\bf #1}}

\begin{document}
\title{Deuteron Matrix Elements in Chiral Effective Theory at Leading Order}
\author{L. Platter$^1$}\email{lplatter@phy.ohiou.edu}
\author{D.~R. Phillips$^{1,2}$}\email{phillips@phy.ohiou.edu}
\affiliation{$^1$Department of Physics and Astronomy, Ohio University,
Athens, OH 45701, USA\\
$^2$Department of Natural Sciences, Chubu University,
1200 Matsumoto-cho, Kasugai, Aichi 487-8501, Japan}
\date{\today}
\begin{abstract}
We consider matrix elements of two-nucleon operators that arise in chiral
effective theories of the two-nucleon system. Generically,
the short-distance piece of these operators scales as $1/r^n$, with $r$
the relative separation of the two nucleons. We show that, when
evaluated between the leading-order wave functions obtained in this
effective theory, these two-nucleon operators are independent of the
cutoff used to renormalize the two-body problem for $n=1$ and $2$.
However, for $n \geq 3$ general arguments about the short-distance
behavior of the leading-order deuteron wave function show that the
matrix element will diverge.
\end{abstract}
\maketitle

\noindent{\it Introduction:} In recent years chiral effective theory
($\chi$ET) has achieved significant prominence as a technique via
which model-independent results can be obtained for few-nucleon
systems (see Refs.~\cite{Be00,BvK02} for reviews).  $\chi$ET is based
on the realization that the interaction between pions and nucleons is
governed by the approximate chiral symmetry of QCD. Therefore the use
of heavy-baryon chiral perturbation theory, together with Weinberg's
proposal that in few-nucleon systems the quantity that has a
well-behaved chiral expansion is the nucleon-nucleon potential
$V$~\cite{We90,We91}, seems to facilitate a systematic calculation
with which nucleon-nucleon scattering can be well described---as long
as the collision energies are significantly below the chiral-symmetry
breaking scale $\Lambda_\chi$~\cite{bira,evgeny,EM}.  In this approach
$V$---and also by extension two-nucleon-irreducible operators for other
processes---is expanded as a chiral series in the usual chiral
perturbation theory expansion parameter
\begin{equation}
P \equiv \frac{p,m_\pi}{\Lambda_\chi, M},
\end{equation}
where $m_\pi$ is the pion mass, $p$ the momenta of the nucleons
involved in the scattering, and $M$ the nucleon mass.
Although questions have been raised about the consistency of such a
power counting~\cite{Ka96,Be01,ES02,No05} this approach has had
considerable success in describing the scattering data in the two- and
three-nucleon sector.

The theory has also been shown to be consistent---in the
renormalization sense---in the ${}^3$S$_1$--${}^3$D$_1$ channel, the
channel where the two-nucleon bound state deuterium
occurs~\cite{Be01,PVRA05}.  The leading-order wave function of
deuterium, $|\psi_{\rm LO} \rangle$ can therefore be obtained by
solving the Schr\"odinger equation---in either momentum-space or
configuration-space---for two nucleons interacting via the piece of
the $NN$ potential which is of chiral order zero. As first realized by
Weinberg~\cite{We90,We91} that piece consists of the venerable
one-pion-exchange potential, together with a (momentum-independent)
four-nucleon contact interaction. This potential is singular: the
Hamiltonian it generates is unbounded from below. It therefore
requires regularization and renormalization. In practice, the
potential is regulated at some momentum scale $\Lambda$ (or in
co-ordinate space at a distance $1/\Lambda$). The strength of the
contact interaction is then adjusted to reproduce some observable,
usually the deuteron binding energy. If this can be done over a wide
range of $\Lambda$, and if other $NN$ scattering observables are
independent of $\Lambda$ up to corrections of higher order in the
$\chi$ET, then we say that the potential has been renormalized. In
Ref.~\cite{Be01} Beane and collaborators showed that this could be
done for the ${}^3$S$_1$--${}^3$D$_1$ channel---at least if one was
only concerned about the results in the chiral limit $m_\pi=0$. This
conclusion has since also been reached using a different regulator and
in a regulator-independent fashion in Refs.~\cite{No05,PVRA05,Bi05}.

With these wave functions in hand it is natural to re-examine the many
successes that phenomenological potential models have had in
describing the deuteron's interaction with external
probes, such as electrons, pions, or photons.  In such a calculation
we use an operator $\hat{O}$ that is appropriate to the particular
external probe under consideration, and has been derived using a
chiral expansion in powers of $P$. That operator is then sandwiched
between a wave
function obtained from a chiral potential (which presumably should be
computed to the same relative order), yielding matrix elements:
\begin{equation}
{\cal M}=\langle \psi|\hat{O}|\psi \rangle.
\label{eq:M}
\end{equation}
Many such calculations have been performed (see Ref.~\cite{Ph05} for a
recent, partial summary), although, with one exception which we will
discuss below, none of them have employed the wave functions of
Refs.~\cite{Be01,No05,PVRA05}. In these calculations the bound-state
wave function is found by employing some regularization prescription
involving a cutoff $\Lambda$.  Here we will discuss the conditions
under which the use of such wave functions in the evaluation of
Eq.~(\ref{eq:M}) yields a matrix element that is independent of
$\Lambda$. The extent of the $\Lambda$ dependence in the result for
${\cal M}$ tells us the degree to which the $\chi$ET prediction for
the matrix element is model independent.

\begin{figure}[t]
\centerline{\includegraphics*[width=3.2in,angle=0]{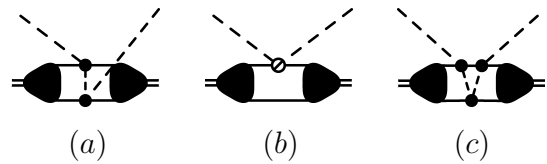}}
\caption{\label{fig:1} Three Feynman diagrams for different contributions to 
the $\pi$d scattering length in chiral effective theory. The dots are vertices from ${\cal L}_{\pi N}^{(1)}$, while the sliced blob is a vertex from 
${\cal L}_{\pi N}^{(2)}$.}
\end{figure}

Recently, Mei\ss ner {\it et al.}~\cite{Me05} and Nogga and Hanhart
\cite{Nogga:2005fv} considered this question for pion-deuteron
scattering, a reaction originally discussed within this framework by
Weinberg in 1992~\cite{Weinberg:1992yk}.  There a two-nucleon operator
representing the process depicted in Fig.~\ref{fig:1}(a) gives a large
contribution to $a_{\pi d}$.  The presence of this large
double-scattering term can obscure attempts to extract the isoscalar
pion-nucleon scattering length from $a_{\pi d}$.  Pion scattering on the
individual nucleons in the deuterium nucleus (see Fig.~\ref{fig:1}(b))
is $O(P^2)$ in the $\chi$ET, and yields a piece of the $\pi$d scattering
length~\cite{EW,Weinberg:1992yk}:
\begin{equation}
a_{\pi d}^{(b)}=\frac{(1+\mu)}{(1+\mu/2)}(a_{\pi n} + a_{\pi
  p}),
\label{eq:1B}
\end{equation}
with $\mu=m_\pi/M$. Meanwhile the double-scattering diagram of
Fig.~\ref{fig:1}(a) gives~\cite{EW,Weinberg:1992yk,Be98}:
\begin{equation}
a_{\pi d}^{(a)}=-\frac{1}{4 \pi^2 (1+\mu /2)}\left(\frac{m_\pi}{2 f_\pi^2}\right)^2
\Bigg\langle\frac{1}{r}\Bigg\rangle,
\label{eq:NLO}
\end{equation}
where here, and throughout what follows, 
\begin{equation}
\langle f(r) \rangle \equiv \int_0^\infty \hbox{d}r \, f(r) (u^2(r) +
w^2(r)), 
\end{equation}
with $u$ and $w$ the ${}^3$S$_1$ and ${}^3$D$_1$ radial deuteron wave
functions. 

The contribution of Eq.~(\ref{eq:NLO}) to the pion-deuteron scattering
length is suppressed by one power of $P$ relative to the
nominally-leading contribution (\ref{eq:1B}). However, since
corrections to the leading-order $NN$ potential are suppressed by two
powers of $P$~\cite{bira}, a next-to-leading order computation of
$a_{\pi d}$ involves the evaluation of the matrix element:
\begin{equation}
\Bigg\langle\frac{1}{r}\Bigg\rangle_{{\rm LO}, \Lambda},
\label{eq:me1}
\end{equation}
where the subscript ${\rm LO}$ indicates that the matrix element
should be taken with deuteron wave functions obtained with the
leading-order $\chi$PT $NN$ potential, and the subscript $\Lambda$ is
included in Eq.~(\ref{eq:me1}) because a cutoff must be applied to that
potential before it is renormalized to give the correct
deuteron binding energy. Since there are no short-distance pieces of
the $\pi$-d operator $\hat{O}$ at next-to-leading order the matrix
element (\ref{eq:me1}) must be cutoff independent over a significant
$\Lambda$ range if Weinberg's approach is to be a consistent way to
calculate pion-deuteron scattering.

Nogga and Hanhart gave numerical evidence that the matrix element
(\ref{eq:me1}) has a $\Lambda \rightarrow \infty$ limit that exists
and is finite. (This is in accord with a similar conclusion previously
obtained in Ref.~\cite{PVRA05}, where $\langle \frac{1}{r} \rangle$
was evaluated using wave functions derived solely from one-pion
exchange. The details of the evaluation of Ref.~\cite{PVRA05} will be
explained further below.)  This shows, Nogga and Hanhart claim, that
expanding $\hat{O}$ in powers of $P$ is a consistent way to calculate
the pion-deuteron scattering length. Consequently, at least in principle, it
should be possible to perform a high-accuracy, model-independent
extraction of the pion-neutron scattering length from pion-deuteron
scattering data.

But other two-nucleon operators that occur in $\pi$d scattering
computations have higher powers of $r$ in the denominator in their
co-ordinate space form. For instance, the next term in the $\pi$N
multiple-scattering series, depicted in Fig.~\ref{fig:1}(c), yields a
contribution to $a_{\pi d}$~\cite{Be02}:
\begin{equation}
a^{(c)}_{\pi d}= \frac{1}{16 \pi^3 (1+\mu /2)}
\left(\frac{m_\pi}{2 f_\pi^2}\right)^3
\Bigg \langle \frac{1}{r^2} \Bigg \rangle~.
\label{eq:O3}
\end{equation}
This contribution is of order $P^5$, and so is N$^3$LO in
the chiral expansion for $a_{\pi d}$, but numerically it is the next two-nucleon
effect that must be considered after the inclusion of the matrix
element (\ref{eq:me1})~\cite{Be02}. The failure of the standard chiral
expansion to account for the large size of such effects is related to
the absence of any suppression of the operator (\ref{eq:O3}) at long
distances. It led the authors of Ref.~\cite{Be02} to propose a
different power counting for the $\pi$d scattering length---one that
more accurately captures the relative hierarchy of mechanisms
contributing to $a_{\pi d}$. Regardless of what the correct counting
for the operator (\ref{eq:O3}) is, if the impact of the
triple-scattering mechanism of Fig.~\ref{fig:1}(c) on extractions of
the isoscalar pion-nucleon scattering length from $a_{\pi d}$ is to be
assessed we must not only consider the matrix element (\ref{eq:me1}),
but also:
\begin{equation}
\Bigg \langle \frac{1}{r^2} \Bigg \rangle_{{\rm LO}, \Lambda}.
\end{equation}

More generally, all two-nucleon operators in the chiral effective theory
will scale as $1/r^n$ at distances $\ll 1/m_\pi$ and $\ll 1/q$ (with $q$
the magnitude of any momentum or energy transferred to the nucleus by
the probe). The value of $n$ is determined by the process under
consideration and the order to which the two-body operator is
calculated, with higher values of $n$ being reached as higher orders
are computed in the chiral expansion for $\hat{O}$. In a perturbative
calculation of deuteron matrix elements within $\chi$ET
the highest-order (and most singular) pieces of the operator
will contribute to the overall result for ${\cal M}$ through their expectation value
taken with the leading-order wave function.  Thus, we are led to a very
general question: If one considers the expression
\begin{equation}
\Bigg \langle \frac{1}{r^n}  \Bigg \rangle_{{\rm LO}, \Lambda}~,
\label{eq:me2}
\end{equation}
then for which values of $n$ does the $\Lambda \rightarrow \infty$ 
limit exist, and for which is it finite? It is this question, together
with associated ones involving operators that connect the S- and
D-wave components of the deuteron wave function, that  we
will answer in this paper.\\ \\
{\it Theory:} Nogga and Hanhart solved the 
Schr\"odinger equation in its momentum-space 
form, i.e. the homogeneous Lippmann-Schwinger equation:
\beq
\label{eq:schroedinger}
\langle {\bf p}|\psi_{\rm LO} \rangle_\Lambda=
G_0(p)\int_0^\Lambda \frac{\hbox{d}^3 p'}{(2 \pi)^3} \,
V^{(0)}(\vec{p},\vec{p}') \langle {\bf p}'|\psi_{\rm LO} \rangle_\Lambda~,
\eeq 
where
$\Lambda$ is the scale at which the potential $V$ is regulated, and
$G_0(p)=(-B_d-p^2/M)^{-1}$ is the (free, center-of-mass frame)
two-nucleon propagator, with $B_d$ and $M$ denoting the deuteron
binding energy and nucleon mass, respectively. The leading-order
potential is given by a one-pion exchange (OPE) contribution and a
short-distance piece: 
\begin{eqnarray}
V^{(0)}({\vec{q}})&=&-\left(\frac{g_A}{2
  f_\pi}\right)^2 {\bf \tau_1 \cdot \tau_2} \frac{(\vec{\sigma_1}\cdot
  \vec{q}) (\vec{\sigma_2}\cdot\vec{q})}{\vec{q}^2+m_\pi^2}
\nonumber + \frac{1}{4\pi} C_t P_t,\nonumber\\
\label{eq:LOV}
\end{eqnarray} 
with $\vec{q} \equiv \vec{p}' - \vec{p}$ the three-momentum of
the exchanged pion. In Eq.~(\ref{eq:LOV}) $P_t$ is a projection
operator that projects onto the ${}^3$S$_1$ channel, and $C_t$ the
strength of the short-distance potential in that channel. While the
OPE contribution is totally determined at leading order through the
pion mass $m_\pi$, the axial coupling constant $g_A=1.26$, and the
pion-decay constant $f_{\pi}=92.4$~MeV, the contact interaction
parameter $C_t$ has to be determined from $NN$ data and will be a
function of the cutoff $\Lambda$.

From Eqs.~(\ref{eq:schroedinger}) and (\ref{eq:LOV}) it is
straightforward to obtain the coupled one-dimensional differential
equations which describe the deuteron wave function:
\bea
\label{eq:deuterondiffeq}
-u''(r)+U_s(r)u(r)+U_{sd}(r) w(r)&=&-\gamma^2 u(r),
\nonumber\\
-w''(r)+U_{sd}(r)u(r)+\biggl[U_d(r)+\frac{6}{r^2}\biggr]w(r)&=&
-\gamma^2 w(r),
\nonumber\\
\label{eq:SErspace}
\eea
where $u$ and $w$ are, as defined above, the deuteron radial wave functions.
The coupled-channel potential is given by
\beq
U_s=U_c,\qquad U_{sd}=2\sqrt{2}U_T,\qquad U_d=U_C-2U_T,
\eeq
with 
\bea
U_C&=&-\frac{m_\pi^2 M g_A^2}{16\pi f_\pi^2}\frac{e^{-m_\pi r}}{r}~,
\nonumber\\
U_T&=&-\frac{m_\pi^2 M g_A^2}{16\pi f_\pi^2}\frac{e^{-m_\pi
    r}}{r}\biggl(1+\frac{3}{m_\pi r}
+\frac{3}{(m_\pi r)^2}\biggr)~.\nonumber\\
\label{eq:LOpotrspace}
\eea
Equations (\ref{eq:SErspace})--(\ref{eq:LOpotrspace})
will be equivalent to Eqs.~(\ref{eq:schroedinger}) and
(\ref{eq:LOV}) provided that $r > 1/\Lambda$.
The equations (\ref{eq:SErspace}) are solved subject to 
the following boundary conditions as $r \rightarrow \infty$:
\bea
u(r)&\rightarrow& A_S \,e^{-\gamma r}~,
\nonumber\\
w(r)&\rightarrow& \eta \,A_S \,e^{-\gamma r}\biggl(1+\frac{3}{\gamma
  r}+\frac{3}{(\gamma r)^2}\biggr)~,
\eea
with $\gamma=\sqrt{M B_d}$ the deuteron wave number, $A_S$ the
normalization constant which guarantees that
\bea
\int\hbox{d}r\bigl( u^2(r) +w^2(r)\bigr)=1~,
\eea
and $\eta$ the asymptotic D/S ratio.

By employing Eqs.~(\ref{eq:SErspace})--(\ref{eq:LOpotrspace}), and
solving them for arbitrarily short distances $r$, Pavon Valderrama and
Ruiz Arriola have calculated $u$ and $w$ using boundary conditions at
inter-nucleon distances of order 0.1--0.2 fm~\cite{PVRA05}. This echoes
the much earlier work of Sprung and collaborators~\cite{Sp00}, as well
as the solution of the leading-order potential with a square-well
regulating the short-distance behavior~\cite{Be01}.  Within such an
approach, an analysis of the asymptotic short-distance behavior of the
components $u$ and $w$ gives the following
result~\cite{Be01,PVRA05,Sp00}: \bea
\label{eq:asymptoticdeuteron}
u_{sd}(r)&=&A_S \frac{1}{\sqrt{3}}
\biggl(\frac{r}{R}\biggr)^{3/4}\biggl[-C_{2R}e^{-4\sqrt{2}\sqrt{R/r}}
  \nonumber\\
  &&\qquad\qquad+2^{3/2}|C_{2A}|\cos\left(4\sqrt{\frac{R}{r}}+\phi\right)\biggr],
\nonumber\\ w_{sd}(r)&=&A_S\frac{1}{\sqrt{3}}
\biggl(\frac{r}{R}\biggr)^{3/4}\biggl[\sqrt{2}C_{2R}e^{-4\sqrt{2}\sqrt{R/r}}
  \nonumber\\
  &&\qquad\qquad+2|C_{2A}|\cos\left(4\sqrt{\frac{R}{r}}+\phi\right)\biggr].
\eea $C_{2A}$ and $C_{2R}$ are normalization constants which have been
determined in \cite{PVRA05}, $R$ is a new length scale that enters the
non-perturbative problem. It is defined
by $R={\textstyle \frac{3 g_A^2 M}{32 \pi f_\pi^2}}$. When 
Eqs.~(\ref{eq:SErspace})--(\ref{eq:LOpotrspace}) are solved in
Ref.~\cite{PVRA05} the phase $\phi$ is determined by the boundary
condition at $r=0$, and so $\phi$ is regulator independent, and is a
function only 
of the scales $m_\pi$, $\gamma$, and $R$.
After computing the numerical solution of
Eqs.~(\ref{eq:deuterondiffeq}) to a sufficiently small radius $r$, it
can be matched to the $r \rightarrow 0$ form of the deuteron wave
function (\ref{eq:asymptoticdeuteron}) and an---in
principle---regulator-independent wave function can be
obtained~\footnote{Although the final wave function is in principle
  regulator independent, the boundary conditions employed in the
  numerical computation of the long-range components can lead to
  cutoff effects at short distances and this leads to a small
  numerical uncertainty of the constants $C_{2A}$, $C_{2R}$ quoted
  in~\cite{PVRA05}. However, with sufficient care this uncertainty can
  be made arbitrarily small.}.\\ \\
\noindent{\it Results:}
Let us now  return to the problem of computing
\begin{equation}
\Bigg \langle \frac{1}{r^n} \Bigg \rangle_\Lambda.
\end{equation}
First, we split the integral up as:
\begin{eqnarray}
&&\Bigg \langle \frac{1}{r^n} \Bigg \rangle_\Lambda
=\int_0^\frac{1}{\Lambda} \hbox{d}r \, \frac{u^2(r) + w^2(r)}{r^n}\nonumber\\
&& \qquad +\int_\frac{1}{\Lambda}^{R*} \hbox{d}r \, \frac{u^2(r) + w^2(r)}{r^n}
\nonumber
+ \int_{R*}^\infty\frac{u^2(r) + w^2(r)}{r^n}~,\nonumber\\
\label{eq:split}
\end{eqnarray}
where $R*$ is sufficiently small that the asymptotic forms
(\ref{eq:asymptoticdeuteron}) apply, but is still $\gg 1/\Lambda$.
The piece of the integral from $R*$ to $\infty$ can be calculated
within $\chi$ET and will depend only on low-energy scales such as
$\gamma$ and $m_\pi$, and, of course, on $R*$ itself.  Meanwhile, as
$\Lambda \rightarrow \infty$ the first integral goes to zero, as long
as the integrand is integrable. Note that any regulator dependence of
the wave functions should be contained in that first piece of our integral,
since we have already explained that the wave functions for $r >
1/\Lambda$ are regulator-independent solutions of
Eqs.~(\ref{eq:SErspace})--(\ref{eq:LOpotrspace}).
Therefore, in order to establish whether or not $\langle
1/r^n \rangle_\Lambda$ is regulator-independent 
the pertinent piece of Eq.~(\ref{eq:split}) is
\begin{equation}
\Bigg \langle \frac{1}{r^n} \Bigg \rangle_{sd}
\equiv \int_\frac{1}{\Lambda}^{R*} \hbox{d}r \, \frac{u^2(r) +
  w^2(r)}{r^n}.
\label{eq:me3}
\end{equation}
If this has a $\Lambda \rightarrow \infty$ limit that is finite then
the entire matrix element $\langle 1/r^n \rangle_\Lambda$ will
also be well behaved in that limit. Inside the integral
Eq.~(\ref{eq:me2}) we may substitute the expressions
Eq.(\ref{eq:asymptoticdeuteron}) for $u$ and $w$. We then see that
short-distance integrals involving the exponential piece of the wave
function will always converge for any $n$, as the exponential itself
regularizes the result.  On the other hand, the term including the
cosine function does not vanish at the origin and therefore we have to
determine the values of $n$ for which the integral
\beq 
\int_{1/\Lambda}^{R*} \hbox{d}r\,r^{3/2-n}
\cos^2\left(4\sqrt{\frac{R}{r}}+\phi\right)
\label{eq:me4}
\eeq 
gives a finite result as $\Lambda \rightarrow \infty$.

This question is easily answered by simple
dimensional analysis. For $n=3$ or higher the integral in
Eq.~(\ref{eq:me4}) will diverge. However, for $n=1,2$ if we ignore
the exponential pieces of $u$ and $w$ we can solve the
integrals analytically and obtain
\begin{eqnarray}
\label{eq:sd1overrn}
&& \Bigg \langle \frac{1}{r^n} \Bigg \rangle_{sd}=
\frac{4 A_S^2 |C_{2A}|^2}{R^{3/2}} \left[\frac{1}{5-2n}\left\{R_*^{5/2-n} -
  \frac{1}{\Lambda^{5/2-n}}\right\}\right.\nonumber\\
&&\qquad + \left. 8^{5-2n} R^{5/2-n} \left\{f_n\left(8\sqrt{R \Lambda}\right) - f_n\left(8 \sqrt{\frac{R}{R_*}}\right)\right\}\right],\nonumber\\
\end{eqnarray}
where:
\begin{equation}
f_n(x) \equiv \int_a^x \hbox{d}t \, t^{2n-6} \cos(t+2\phi).
\label{eq:fn}
\end{equation}
The function $f_n$ can be written as a linear combination of
incomplete Gamma functions. Its asymptotic expansion is:
\begin{equation}
f_n(x) \rightarrow \alpha_n + h_n(a) + x^{2n-6} \sin(x+ 2 \phi) + O(x^{2n-7}),
\end{equation}
as $x \rightarrow \infty$ ($n=1,2$), with:
\begin{equation}
\alpha_1=\frac{\pi}{12} \cos(2 \phi); \qquad 
\alpha_2=-\frac{\pi}{2} \cos(2\phi).
\end{equation}
The function $h_n(a)$ can easily be evaluated, but its value is not
relevant for our purposes here, since it cancels in Eq.~(\ref{eq:sd1overrn}).
Therefore for $n=1$ and $2$ the matrix element has 
a finite $\Lambda \rightarrow \infty$ limit. However, that limit is
approached in the presence of cutoff-dependent oscillations whose
amplitude scales as $(R \Lambda)^{n-3}$.  Another cycle in this oscillation
occurs whenever $1/\Lambda$ becomes small enough that we integrate over
another node in the deuteron wave function. The presence of
oscillations is thus associated with the appearance of spurious bound
states in the effective theory---bound states whose binding energies
are greater than the theory's breakdown scale.  However,
for the $n=1$ and $n=2$ matrix elements, the oscillations vanish as $\Lambda
\rightarrow \infty$, so even though such bound states are present, the
procedure of evaluating the matrix element with a wave function at a
given $\Lambda$ should yield the correct (i.e. $\Lambda \rightarrow
\infty$) answer, as long as a sufficiently high cutoff is chosen.

Now adding the $\Lambda \rightarrow \infty$ piece of these matrix elements
to the piece that comes from integration between $R*$ and infinity,
we find: 
\bea \Bigg
\langle\frac{1}{r} \Bigg \rangle_{\rm LO}&=& 0.478~\hbox{fm}^{-1},
\label{eq:1overr}\\
\Biggl\langle\frac{1}{r^2}\Biggr\rangle_{\rm LO}&=&0.425~\hbox{fm}^{-2},\label{eq:1overr2}
\eea
where the first result is in agreement with the number given in
Ref.~\cite{PVRA05}. As pointed out there, it is essentially consistent
with the range $0.450$--$0.465$ obtained for $\langle
1/r \rangle$ using a variety of potential-model wave
functions in Ref.~\cite{Be02}. 
The result (\ref{eq:1overr2}) is new, and does not fall within
the range $0.286$--$0.345$ fm$^2$ quoted in
Ref.~\cite{Be02}. It takes longer to reach asymptopia for this matrix element.
\begin{figure}[tb]
\centerline{\includegraphics*[width=2.8in,angle=0]{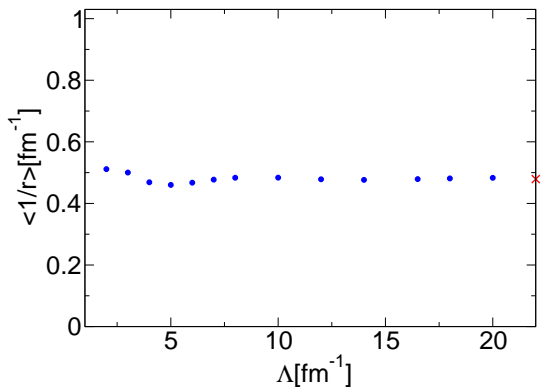}}
\caption{\label{fig:r1}Cutoff-dependence of the matrix element
  $\langle {\textstyle\frac{1}{r}}\rangle_\Lambda$, using the radial
  deuteron wave functions employed in \cite{Nogga:2005fv}. The cross
  is the result found using the wave function of Ref.~\cite{PVRA05},
  as given in Eq.~(\ref{eq:1overr}).}
\end{figure}

\begin{figure}[tb]
\centerline{\includegraphics*[width=2.8in,angle=0]{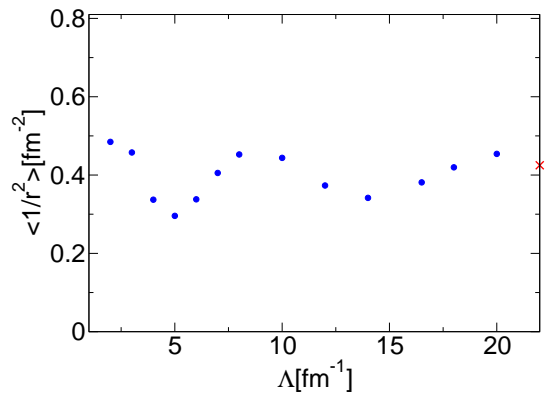}}
\caption{\label{fig:r2}Cutoff-dependence of $\langle
  {\textstyle\frac{1}{r^2}}\rangle_\Lambda$, using the
radial deuteron wave functions employed in \cite{Nogga:2005fv}. The
cross is the result found using the wave function of
Ref.~\cite{PVRA05}, as given in Eq.~(\ref{eq:1overr2}).}
\end{figure}

In order to check the results of Eqs.~(\ref{eq:1overr}) and
(\ref{eq:1overr2}) we have computed the expectation value $\langle
\psi_{\hbox{{\tiny LO}}}|{\textstyle\frac{1}{r^n}}|\psi_{\hbox{{\tiny
      LO}}}\rangle$ for $n=1$ and $2$. We used the same wave functions
as were used in \cite{Nogga:2005fv} with cutoff values between 2 and
20 fm$^{-1}$.  As Fig.~\ref{fig:r1} shows, we find similar behavior
for $\langle 1/r \rangle$ as was displayed in that paper, and a
limiting value that is consistent with that found by Nogga and Hanhart~\cite{No06}.
But, we can now interpret the $\Lambda$-dependent oscillations seen in
Fig.~\ref{fig:r1} as exactly the ones predicted by
Eq.~(\ref{eq:sd1overrn}).  Meanwhile, Fig.~\ref{fig:r2} shows that,
as expected based on our analysis of Eq.~(\ref{eq:sd1overrn}), these
oscillations are more pronounced for $\langle 1/r^2 \rangle$, and in
consequence a much higher cutoff is needed to achieve a converged
result. However, also for this matrix element the numerical results
indicate that a finite limiting value exists, which is in accord with
our previously presented analytic arguments.

Since Eq.~(\ref{eq:asymptoticdeuteron}) identifies the relevant
behavior of the $r \rightarrow 0$ piece of the deuteron wave function,
we are able to predict that all powers $n \geq 3$ will lead to
divergent results. This result is supported by numerical calculations
of the corresponding matrix elements with the wave functions used in
\cite{Nogga:2005fv}.

We also find that, to a very good approximation, the result for
\begin{equation}
\lim_{\Lambda \rightarrow \infty} \Bigg \langle \frac{1}{r^n}
\Bigg \rangle_{{\rm LO}, \Lambda}
\end{equation}
agrees (for $n=1$ and $2$) with the numerical results obtained with
the co-ordinate space wave function of Ref.~\cite{PVRA05} that we
presented in Eqs.~(\ref{eq:1overr}) and (\ref{eq:1overr2}).  (The
slight disagreement exhibited in Figs.~\ref{fig:r1} and \ref{fig:r2}
between the trend of the dots and the cross that represents the result
of Eqs.~(\ref{eq:1overr}) and (\ref{eq:1overr2}) can presumably be
traced to the 3\% difference in the $\pi$NN coupling constants used in
Refs.~\cite{PVRA05} and \cite{Nogga:2005fv}.) This is an important
result, because, together with the even better agreement for matrix
elements of positive powers of $r$ (see Table I of
Refs.~\cite{PVRA05,Nogga:2005fv}) it suggests that the procedure of
constructing a limiting sequence of cutoff-dependent wave functions
and evaluating the matrix element as a function of $\Lambda$ is not
actually necessary. Instead such matrix elements can be computed using
the co-ordinate space results for the deuteron wave function given in
Refs.~\cite{Sp00,Be01,PVRA05}.

\begin{figure}[tb]
\centerline{\includegraphics*[width=2.8in,angle=0]{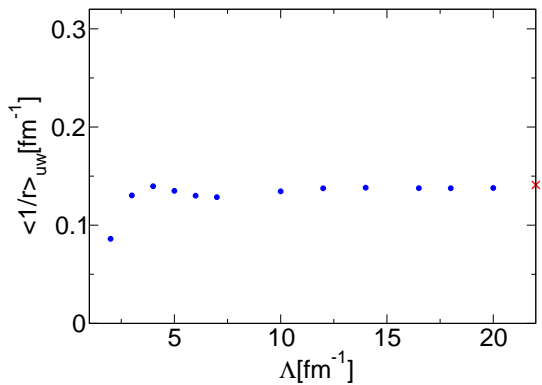}}
\caption{\label{fig:uwint1}Cutoff-dependence of the integral $\int
  \hbox{d}r\,u_\Lambda(r) w_\Lambda(r)/r$, where $u_\Lambda$
  ($w_\Lambda$) is the ${}^3$S$_1$ (${}^3$D$_1$) radial deuteron wave
  function found for cutoff $\Lambda$ in Ref.~\cite{Nogga:2005fv}. The
  cross indicates the result of evaluation with the wave functions of
  Ref.~\cite{PVRA05}, Eq.~(\ref{eq:uwoverr}).}
\end{figure}

\begin{figure}[tb]
\centerline{\includegraphics*[width=2.8in,angle=0]{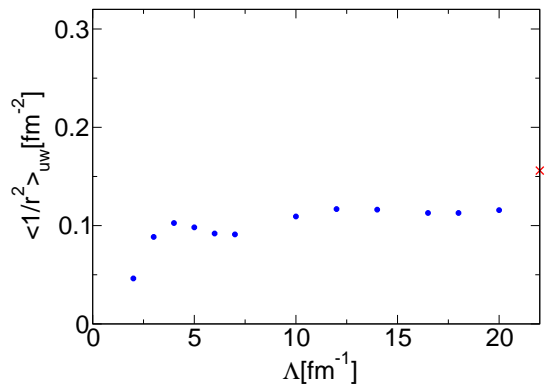}}
\caption{\label{fig:uwint2} As for Fig.~\ref{fig:uwint1}, but for
$\int \hbox{d}r\,u_\Lambda(r) w_\Lambda(r)/r^2$. In this instance the
  cross is the result of Eq.~(\ref{eq:uwoverr2}).}
\end{figure}

Next we consider integrals of the form:
\bea
\int\hbox{d}r \frac{u(r)w(r)}{r^n}~.
\label{eq:uwint}
\eea
For $n=1$ this integral occurs in other, numerically less important,
corrections to the $\pi$d scattering length. For $n=2$ contributions
of the form (\ref{eq:uwint})
occur in the evaluation of sub-leading pieces of two-nucleon effects in
the process $\gamma$d$\rightarrow \pi^0$d.
Since $u$ and $w$ both display the same short-distance behavior, 
we arrive at the same conclusion as for the matrix
elements considered above: for $n=1$ and $n=2$ the result is
convergent (in spite of oscillations), see Figs.~\ref{fig:uwint1} and
\ref{fig:uwint2}, while for $n \geq 3$
it is divergent. For $n=1$ and $2$ evaluation with the wave functions
of Ref.~\cite{PVRA05} gives:
\begin{eqnarray}
\int_0^\infty \hbox{d}r \frac{u(r)w(r)}{r}&=&0.141~\hbox{fm}^{-1},
\label{eq:uwoverr}\\
\int_0^\infty \hbox{d}r \frac{u(r)w(r)}{r^2}&=&0.156~\hbox{fm}^{-2}.
\label{eq:uwoverr2}
\end{eqnarray}
Both of these numbers are consistent with the trend of the results for
finite $\Lambda$, as shown in Figs.~\ref{fig:uwint1} and
\ref{fig:uwint2}.  However, the second result will not be attained
until $\Lambda \gg 20~{\rm fm}^{-1}$, since approximately 20\% of the
final number accrues in the region between $r=0$ and $r=0.1$~fm.\\
\\
\noindent{\it Conclusion:} In this paper we have shown explicitly that
deuteron matrix elements of two-nucleon operators which are
proportional to $1/r^n$ at short distances converge for $n \leq 2$ and
diverge for $n\geq 3$ when they are evaluated using the leading-order
deuteron wave function. We have given numerical evidence by explicit
calculations using a sequence of leading-order deuteron wave functions
corresponding to different ultraviolet cutoffs, and the results
obtained in this way agree with an analysis based on the
short-distance behavior of the deuteron wave function.

As higher orders are calculated in the chiral series for the operator
$\hat{O}$ the divergent case $n=3$ will be reached
(e.g. $\gamma$d$\rightarrow \pi^0$d at $O(P^4)$ in
$\chi$PT~\cite{Be97}). Suppose this occurs at some chiral order
$m$. If both the wave function $|\psi \rangle$ and the operator
$\hat{O}$ are written as a chiral series, then this implies that an
$m$th-order piece of the matrix element (\ref{eq:M}), specifically the
piece $\langle \psi_{\rm LO}|\hat{O}^{(m)}|\psi_{\rm LO} \rangle$, is
divergent. This apparently mandates the presence of a contact
interaction involving both nucleons and the external probe in
$\hat{O}^{(m)}$, so that this divergence can be absorbed. However, in
explicit calculations of a particular process to order $m$ in the
chiral expansion this divergence may cancel with contributions to
(\ref{eq:M}) due to corrections to the leading-order wave
function. Such corrections will come from higher-order pieces in the
chiral expansion of the $NN$ potential, and will generate
contributions to (\ref{eq:M}), e.~g., of the type $\delta \langle
\psi^{(m)}|\hat{O}^{\rm LO}|\psi_{\rm LO} \rangle$. Therefore, the
appearance of a matrix element of an operator $1/r^n$ with $n \geq 3$
at order $m$ does not immediately indicate that a counterterm is
needed at that order. However, if a counterterm is {\it not} permitted
at that order by chiral symmetry and/or electromagnetic gauge
invariance, then the various divergences at order $m$ must cancel each
other. This represents a constraint on the sum of all mechanisms that
contribute to (\ref{eq:M}) at that order.  (For an example of such a
cancellation in the context of pion production, see
Refs.~\cite{Ga05,Le05}.)

Lastly, we note that if an $NN$ potential that is more singular than
one-pion exchange is iterated to all orders using the Schr\"odinger
equation, as is done in Refs.~\cite{bira,evgeny,EM}, then the behavior of
the wave functions $u$ and $w$ will not follow the form
(\ref{eq:asymptoticdeuteron}). Indeed, the more singular the
potential, the more convergent $u$ and $w$ will become~\cite{PVRA06}.
Therefore, the conclusion of the previous paragraph is limited to an
approach to higher-order calculations in the chiral effective theory
where all corrections to the leading-order $|\psi \rangle$ are
evaluated in perturbation theory, and so the form
(\ref{eq:asymptoticdeuteron}) still represents the dominant
short-distance behavior of $u$ and $w$.\\ 
\\
\noindent{\it Acknowledgments:} We thank M.~Pavon Valderrama and
E.~Ruiz Arriola for discussions and for supplying us with numerical
results for the deuteron wave function. We are also grateful to
A.~Nogga for sending us co-ordinate space forms of the wave functions
of Ref.~\cite{No05}, and for comments on the manuscript. D.~R.~P.
thanks Silas Beane and Martin Savage for conversations that initiated
this study. L.~P. thanks Chubu University for its hospitality during
the completion of this work. We acknowledge support by the
U.S. Department of Energy under grant DE-FG02-93ER40756.

\end{document}